# Self-homodyne measurement of a dynamic Mollow triplet in the solid state

Kevin A. Fischer[1†★], Kai Müller[1†], Armand Rundquist[1], Tomas Sarmiento[1], Alexander Y. Piggott[1], Yousif Kelaita[1], Constantin Dory[1], Konstantinos G. Lagoudakis[1], Jelena Vučković[1★]

**The study of light-matter interaction at the quantum scale has been enabled by the cavity quantum electrodynamics (CQED) architecture,[1] in which a quantum two-level system strongly couples to a single cavity mode. Originally implemented with atoms in optical cavities,[2,3] CQED effects are now also observed with artificial atoms in solid-state environments.[4–6] Such realizations of these systems exhibit fast dynamics, which makes them attractive candidates for devices including modulators and sources in high-throughput communications. However, these systems possess large photon out-coupling rates that obscure any quantum behavior at large excitation powers. Here, we have utilised a self-homodyning[7] interferometric technique that fully employs the complex mode structure of our nanofabricated cavity[8–10] to observe a quantum phenomenon known as the dynamic Mollow triplet.[11] We expect this interference to facilitate the development of arbitrary on-chip quantum state generators, thereby strongly influencing quantum lithography, metrology, and imaging.**

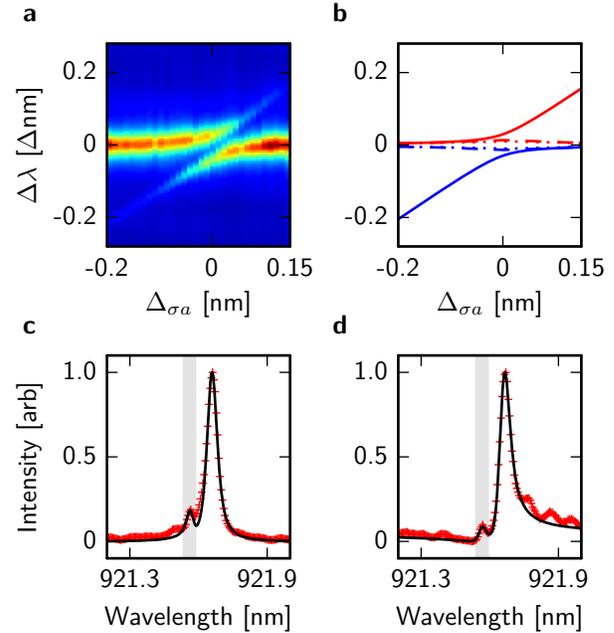

**Figure 1 | Characterization of the strongly-coupled system. a,** Cross-polarised reflectivity (mimicking transmission) spectrum of the coupled QD-cavity system obtained when tuning the QD resonance through the cavity mode. By fitting profiles from these spectra, we extract the cavity energy decay rate $\kappa = 2\pi \cdot 15\,\text{GHz}$ and the coherent coupling rate $g = 2\pi \cdot 11\,\text{GHz}$. **b,** Transient energies for climbing the JC-ladder rung by rung for the first, second and third rung as solid, dashed and dotted lines, respectively. Transitions from upper and lower polaritons are colour coded in red and blue, respectively. **c,d,** Spectra of the coupled QD-cavity system taken at a QD-cavity detuning of $\Delta_{\sigma a} = -85\,\text{pm}$ and an excitation power of roughly $15\,\text{nW/nm}$, showing QD-like polaritonic emission (highlighted by grey box) on top of (**c**) a Lorentzian resonance and (**d**) a Fano resonance. In both cases, spectra are taken by performing broadband cross-polarized reflectivity measurements on the system, but under different interference conditions (with altered focus). Red hashes indicate experimental data, black curves indicate quantum-optical fits.

The crowning achievement of quantum optics has been to develop a complete theory for the phenomenon of resonant light scattering from a quantized matter system. Beyond providing closure to debates over the nature of light, this theory has enabled observations of uniquely quantum spectacles such as photon antibunching,[4,12] indistinguishable quantum interference,[5,13] and the Mollow triplet.[14,15] The addition of nanoscale resonators to the quantum scattering problem has provided a new frontier in our quest to mould the flow of light.[4–6] Here, our reported innovation centres on the investigation of resonant light scattering from a quantum nonlinearity (quantum dot [QD]) strongly coupled to a photonic crystal [PC] cavity. This strong coupling allows for quantum-coherent energy exchange between the resonator's quantized light field and the QD's excitonic field, leading to the formation of light-matter entangled states known as polaritons.[16–18] Evidence for the system's strong coupling is provided from the clean avoided-crossing spectra in Fig. 1a. The relative positions of the emission peaks are determined by the transient energies of the Jaynes-Cummings (JC) ladder[19] (Fig. 1b). As the two polaritonic peaks transition through the avoided-crossing (at the QD-cavity detuning of $\Delta_{\sigma a} = 0$), they exchange character from cavity/QD-like to QD/cavity-like. We now detail our unorthodox application of the full resonance structure in PCs[8–10] to generate a self-homodyne interference[7] that emphasizes the hallmark quantum character of the scattered light.

In explaining landmark experiments from solid-state CQED with PCs,[6,16–18] only the fundamental cavity mode was considered to play a role. Recently, however, classical scattering studies of cross-polarised reflectivity through an L3 PC cavity have suggested that the continuum modes above-the-light-line may play an important role.[8–10] This additional scattering was shown to interfere with the PC's fundamental cavity mode resulting in a class of lineshape known as the Fano resonance.[20] To observe the typical background-free JC lineshape for a detuned strongly coupled system, we slightly defocused the sample to reject the continuum mode scattering[8,9] from the L3 PC. Here, we present the results of a broadband cross-polarised reflectivity (identical to transmission) experiment[18] in Fig. 1c. This measurement is a cut from Fig. 1a at $\Delta_{\sigma a} = -85\,\text{pm}$ and shows a small hump of QD-like polaritonic emission on top of a Lorentzian cavity-like resonance. An excellent fit (black line) is obtained with a quantum-optical JC transmission model[19,21] (see Methods).

[1]E. L. Ginzton Laboratory, Stanford University, Stanford, California 94305, USA
†These authors contributed equally
★email: kevinf@stanford.edu; jela@stanford.edu

Next, we returned the sample focus to again destructively interfere the cavity and continuum channels. We now experimentally show that the inclusion of this continuum-mode scattering can interferometrically reject scattered light from the cavity-like polariton in favour of emission from the QD-like polariton. The resulting lineshape (Fig. 1d) shows QD-like polaritonic emission on top of a Fano resonance. Importantly, by comparing this result to Fig. 1c, we see that emission from the cavity-like polariton at the QD-like polaritonic emission wavelength is significantly suppressed (see Supplementary Figs 2 and 3). Although previous experiments have utilized homodyne interference to create Fano-like lineshapes with QD-PC cavity systems,[22] the interference was not capable of reducing light scattered from the cavity-like polariton in favour of emission from the QD-like polariton. In contrast, our unconventional technique enables the continuum modes to interfere with nearly the exact opposite phase from that of the cavity-mode, leading to an apt description: Fano-induced self-homodyne interference. By analogy to a traditional homodyne measurement, this interference allows for the extraction of hidden signals.

Next, we explore the implications of this suppression at high excitation powers and show that only light from the cavity-like polariton is interferometrically cancelled at the QD-like polariton's emission frequency. To this end, we performed simulations and experiments where we resonantly excited the system at the QD-like polariton's emission frequency with a laser pulse. In a formal quantum-mechanical description of scattering,[23] the spectrum of the free-field mode operator $A(t)$ as measured by an ideal infinite-bandwidth detector is given by

$$S(\omega) = \iint_{\mathbb{R}^2} dt\, d\tau\, e^{-i\omega\tau} \langle A^\dagger(t+\tau) A(t) \rangle \quad (1)$$

integrating over all possible two-time correlations $\langle A^\dagger(t+\tau) A(t) \rangle$.

When computing expectations for combinations of $A(t)$ (see Supplementary information), input-output theory[7] can relate the internal cavity mode operator $a(t)$ to the external field operator by the radiative cavity field coupling rate $\kappa/2$. Hence, for a JC system in the solid state where the QD radiative lifetime plays an insignificant role compared to $\kappa$,[24] spectral decomposition of the cavity mode operator yields the spectrum of the detected light. Therefore, we can compute an unnormalized version of this spec-trum with $A(t) \rightarrow a(t)$ in equation (1). In the inset of Fig. 2a, we present experimental data (without Fano filtering) in agreement with this model: only a single peak at the QD-like polaritonic wavelength is observed. Next, we compute the coherent spectrum[23] of the light scattered from the JC system (Fig. 2b); the coherent spectrum is computed with $\langle A^\dagger(t+\tau)A(t)\rangle \rightarrow \langle a^\dagger(t+\tau)\rangle\langle a(t)\rangle$ in equation (1). Aside from the similarity of Figs 2a and 2b, the two-time correlations in Figs 2d and 2e from which the spectra in Figs 2a and 2b are (respectively) derived, exhibit nearly identical profiles. These strong similarities suggest that the scattered light from our JC system at high powers is mostly coherent and thus predominantly classical in nature.

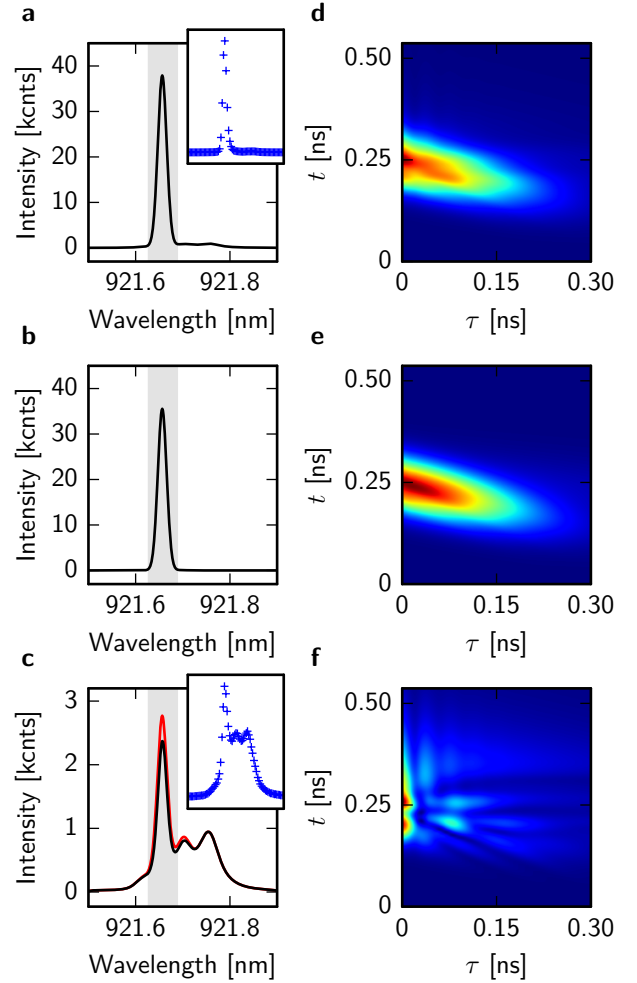

Figure 2 | **Evidence for Fano-induced self-homodyne interference from the detuned strongly-coupled system.** First-order correlations and their spectra under resonant excitation of the QD-like polariton (at the center of the grey box) with an 80 MHz repetition rate pulsed laser of FWHM $\tau_p = 100\,\text{ps}$. **a,** Simulated spectrum of total emission from the JC system; inset, experimental data taken at 500 nW incident power and the interference condition yielding the Lorentzian-like resonance in Fig. 1c (see Supplementary Fig. 1). Note: the inset shares axes with its main panel. **b,** Simulated spectrum of coherently scattered light from the same JC system under identical excitation. **c,** Simulated spectrum of incoherently scattered light (black); simulated spectrum interfering the JC emission with the continuum-mode emission (red), revealing the incoherently scattered light; inset, experimental data taken at 500 nW incident power and the interference condition yielding the Fano-like resonance in Fig. 1d (see Supplementary Fig. 1). Note: the inset shares axes with its main panel. **d-f,** Simulated two-time correlation functions $\langle A^\dagger(t+\tau) A(t) \rangle$ corresponding to their respective spectra in (**a-c**). Plots (**d**) and (**e**) show few differences, but self-homodyne interference reveals the rich correlations in (**f**) due to quantum fluctuations.

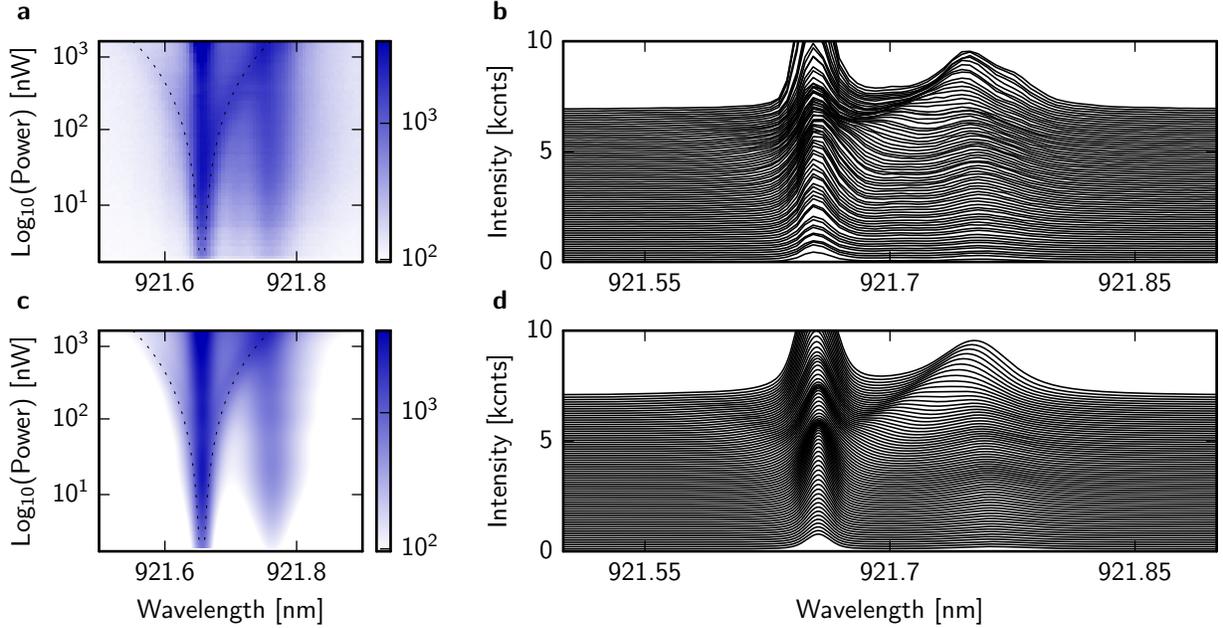

**Figure 3 | Emergence of dynamic Mollow-like triplets from the detuned strongly-coupled system under pulsed resonant excitation of the QD-like polariton. a,b,** Experimental power-series spectra of emission from our system, with self-homodyne interference, as a function of the average power from our $\tau_\mathrm{p} = 100$ ps FWHM pulsed excitation laser. We plot these data both on a log plot (**a**) and as stacked spectra (**b**): the log plot easily shows the power dependence of the side-band frequencies, while the stacked spectra reveal the strong suppression due to the interference. We note that these spectra are taken at the same QD-cavity detuning as the spectra in Fig. 2. Here, two Mollow-like sidebands clearly emerge with increasing excitation power. The presence of the cavity, combined with acoustic phonon scattering, leads to a dramatic asymmetry in the emission from the two side-bands. **c,d,** Simulated power-series spectra, plotted in the same scale and manner as (**a**) and (**b**), showing excellent agreement in both qualitative shape and quantitative values with experimental data in (**a**) and (**b**). Dashed lines show a fit to the power dependence of the triplet frequency splittings (see Supplementary information). Self-homodyne interference clearly reveals and enables the measurement of these unique incoherent scattering spectra from a highly-dissipative strongly coupled system.

On the other hand, the incoherently scattered light reveals the effect of the embedded quantum nonlinearity. The JC incoherent spectrum, the spectrum of cavity-mode operator fluctuations[23] (black line in Fig. 2c), is simulated with

$$\langle A^\dagger(t+\tau)A(t)\rangle \rightarrow \langle a^\dagger(t+\tau)a(t)\rangle - \langle a^\dagger(t+\tau)\rangle\langle a(t)\rangle \quad (2)$$

in equation (1) and clearly shows massive nonlinear conversion at a single photon level as well as rich quantum character in the two-time correlations (Fig. 2f). Strikingly, by careful manipulation of the continuum-mode channel (aforementioned) to enable self-homodyne suppression of the coherently scattered light, we obtain an experimental spectrum (inset of Fig. 2c) remarkably close to the simulated incoherent spectrum. We note that this measurement is extremely phase-stable due to the self-homodyne interference. Note that the left- and right-most peaks occur at the QD-like and cavity-like polaritons (Fig. 1d), respectively; the middle peak will be discussed later.

By comparing spectra in Figs 2a and 2c, we find that we are experimentally capable of rejecting $> 95\,\%$ of the coherently scattered light at this detuning. To model this effect, we replace the free-field operator with a superposition of the cavity mode operator and the scattered coherent state $\alpha(t)$, i.e. $A(t) \rightarrow a(t) + \alpha(t)$ in equation (1) (red line in Fig. 2c). Physically, $\alpha(t)$ is a slightly phase- and amplitude- shifted version of the incident laser pulse (originating from the continuum-mode scattering). For proper choice of $\alpha(t)$, the simulated experimental spectrum is nearly identical to the JC incoherent spectrum. From the excellent agreement between simulation and experiment, we conclude that our Fano-induced self-homodyne measurement leads almost exclusively to the incoherent portion of the strongly coupled system's spectrum.

Armed with the self-homodyne technique, we sought an application to demonstrate its ability to reveal interesting quantum optics: we chose to investigate the theoretically predicted dynamic Mollow triplets that arise under the pulsed excitation of a solid-state system.[11] We emphasize that this effect is quite different from the conventional Mollow triplets (already studied in the solid state[15]) that are observed under strong resonant continuous-wave driving of a two-level system. Specifically, the dynamic Mollow triplets that emerge under pulsed resonant driving of a two-level system have not yet been observed in the solid-state, as they require excitation laser pulse durations longer than the state lifetime with very high powers.[11] Here, we choose the QD-like polariton of our CQED system as our two-level system to be driven because its strong coupling to the cavity provides a solution to these excitation challenges: the cavity coupling decreases the QD state lifetime and the high quality-factor cavity increases polariton coupling to the probe field. Hence, we can observe the formation of dynamic Mollow triplets with driving laser pulses of $\tau_\mathrm{p} = 100\,\mathrm{ps}$ — only approximately two times longer than the QD-like polariton lifetime. Uniquely in the strongly coupled system, as Fig. 1b suggests, some detuning is required in order to generate enough nonlinearity in the JC ladder to drive relatively clear two-photon transitions (see Supplementary information). These cavity enhancements come at the cost of increased coherent scattering that obscures the Mollow triplets, as seen from Fig. 2, and therefore, we are required to employ the self-homodyne technique to reveal this quantum effect.

The striking experimental observation of dynamic Mollow-triplet sidebands is shown in Figs 3a and 3b. The sideband frequencies follow the square root of the probe power, as derived for the detuned CQED system (see Supplementary information). Experimentally, we found the triplets most clear at $\Delta_{\sigma a} = -85\,\mathrm{pm}$. Interestingly, the Mollow-like sidebands are extremely unbalanced: emission at the cavity-like polariton frequency occurs for all powers due to interaction with the higher JC manifolds, which is emphasized by exciton-phonon interaction.[25] These features are shown to be in excellent agreement with a quantum optical simulation of dynamic Mollow triplets from the CQED system with self-homodyne Fano interference (Figs 3c and 3d). Simulations suggest that the maximum cavity occupancy during CQED triplet production is only 1.5–2 photons with variance $< 0.2$, strongly distinguishing the CQED triplets from those arising under the strong-pump limit of a CQED system that leads to the traditional Mollow triplet.[26] Additionally, the strong-coupling-induced asymmetry of the CQED dynamic triplet masks the presence of the additional side peaks theoretically predicted from a two-level system[11] (see Supplementary Fig. S4).

Thus, we have demonstrated a Fano-induced self-homodyne measurement technique and employed it to observe a unique CQED phenomenon in the solid-state — the dynamic Mollow triplet. Hence, we have been able to probe a highly dissipative Jaynes-Cummings system in a high pulse energy regime for the first time, and expect this technique to allow for a more thorough exploration of the rich Jaynes-Cummings dynamics in solid-state systems. The observation of this phenomenon even has practical relevance to the emerging field of Mollow spectroscopy, where the dynamic Mollow triplets may serve as ultrasensitive probes for the nonlinearity of new quantum emitters.[27] Furthermore, we expect the ability to utilize the self-homodyne technique to stably measure quantum fluctuations to find application in the generation of other arbitrary quantum states of light. In particular, we have already shown that our interferometric measurement methodology enables nearly perfect single photon generation in solid-state photon blockade[24] and believe that it will also enable the first generation of indistinguishable photons from an optical solid-state CQED system. Furthermore, the ability to mitigate the unwanted coherent scattering from highly dissipative JC systems should facilitate the direct observation of the JC higher rungs, and thus it may allow for low-detuning photon bundling[28] of arbitrary photon number states, with applications in quantum lithography, metrology, and microscopy. In fact, self-homodyne interference may prove extremely valuable in designing on-chip sources of quantum light, where almost any waveguide-coupled cavity can easily exploit Fano interference.[29] Therefore, in looking toward future applications, our approach may play a pivotal role in generating quantum light while overcoming the inherently strong dissipation of solid-state systems.

## Methods

The MBE-grown structure consists of an $\approx 900\,\mathrm{nm}$ thick $\mathrm{Al_{0.8}Ga_{0.2}As}$ sacrificial layer followed by a 145 nm thick GaAs layer that contains a single layer of InAs QDs. Our growth conditions result in a typical QD density of $(60-80)\,\mu\mathrm{m}^{-2}$. Using 100 keV e-beam lithography with ZEP resist, followed by reactive ion etching and HF removal of the sacrificial layer, we define the photonic crystal cavity. The photonic crystal lattice constant was $a = 246\,\mathrm{nm}$ and the hole radius $r \approx 60\,\mathrm{nm}$. The cavity fabricated is a linear three-hole defect (L3) cavity. To improve the cavity quality factor, holes adjacent to the cavity were shifted.

All optical measurements were performed with a liquid helium flow cryostat at a temperature of $\approx 30\,\mathrm{K}$. A 1.0 m long pulse-shaper was used to generate our experimental excitation pulses from a 3 ps Tsunami mode-locked laser. For excitation and detection a microscope objective with a numeric aperture of $NA = 0.75$ was used. Cross-polarised measurements were performed using a polarising beam splitter. To further enhance the extinction ratio, additional thin film linear polarisers were placed in the excitation/detection pathways and a single-mode fibre was used to spatially filter the detection signal. Furthermore, two waveplates were placed between the beamsplitter and microscope objective: a half-wave plate to rotate the polarisation relative to the cavity and a quarter-wave plate to correct for birefringence of the optics.

Quantum-optical simulations were performed with the Quantum Optics Toolbox in Python (QuTiP),[30] where the standard[21] Jaynes-Cummings model was used as a starting point. For the JC transmission experiments, the cavity occupancy was monitored as the excitation frequency was swept. The scattered coherent state was included through a time-independent operator replacement similar to the one discussed in the main text. The effects of phonons were incorporated through the addition of incoherent decay channels and were subsequently used as fitting parameters. Effective phonon-transfer rates for the $a^\dagger \sigma$ and the $a \sigma^\dagger$ processes were $2\pi \cdot 0.9\,\mathrm{GHz}$ and $2\pi \cdot 0.65\,\mathrm{GHz}$, respectively. These rates are consistent with previous experimental results and a more rigorous analysis.[24]

## References

1. Horoche, S. & Kleppner, D. Cavity quantum electrodynamics. *Phys. Today* **42**, 24 (1989).
2. Nußmann, S. *et al.* Vacuum-stimulated cooling of single atoms in three dimensions. *Nature Physics* **1**, 122–125 (2005).
3. Birnbaum, K. M. *et al.* Photon blockade in an optical cavity with one trapped atom. *Nature* **436**, 87–90 (2005).
4. Michler, P. *et al.* A quantum dot single-photon turnstile device. *Science* **290**, 2282–2285 (2000).

5. Santori, C., Fattal, D., Vucković, J., Solomon, G. S. & Yamamoto, Y. Indistinguishable photons from a single-photon device. *Nature* **419**, 594–597 (2002).
6. Faraon, A. *et al.* Coherent generation of non-classical light on a chip via photon-induced tunnelling and blockade. *Nat. Phys.* **4**, 859–863 (2008).
7. Carmichael, H. J. *Statistical Methods in Quantum Optics* (Springer, 2008).
8. Vasco, J. P., Vinck-Posada, H., Valentim, P. T. & Guimãraes, P. S. S. Modeling of fano resonances in the reflectivity of photonic crystal cavities with finite spot size excitation. *Opt. Express* **21**, 31336–31346 (2013).
9. Galli, M. *et al.* Light scattering and fano resonances in high-q photonic crystal nanocavities. *Appl. Phys. Lett.* **94**, 071101 (2009).
10. Valentim, P. T. *et al.* Asymmetry tuning of fano resonances in GaAs photonic crystal cavities. *Appl. Phys. Lett.* **102**, 111112 (2013).
11. Moelbjerg, A., Kaer, P., Lorke, M. & Mørk, J. Resonance fluorescence from semiconductor quantum dots: beyond the mollow triplet. *Phys. Rev. Lett.* **108**, 017401 (2012).
12. Kimble, H. J., Dagenais, M. & Mandel, L. Photon antibunching in resonance fluorescence. *Phys. Rev. Lett.* **39**, 691–695 (1977).
13. Hong, C. K., Ou, Z. Y. & Mandel, L. Measurement of subpicosecond time intervals between two photons by interference. *Phys. Rev. Lett.* **59**, 2044–2046 (1987).
14. Mollow, B. R. Power spectrum of light scattered by Two-Level systems. *Phys. Rev.* **188**, 1969–1975 (1969).
15. Flagg, E. B. *et al.* Resonantly driven coherent oscillations in a solid-state quantum emitter. *Nat. Phys.* **5**, 203–207 (2009).
16. Yoshie, T. *et al.* Vacuum rabi splitting with a single quantum dot in a photonic crystal nanocavity. *Nature* **432**, 200–203 (2004).
17. Reithmaier, J. P. *et al.* Strong coupling in a single quantum dot-semiconductor microcavity system. *Nature* **432**, 197–200 (2004).
18. Englund, D. *et al.* Controlling cavity reflectivity with a single quantum dot. *Nature* **450**, 857–861 (2007).
19. Laussy, F. P., del Valle, E., Schrapp, M., Laucht, A. & Finley, J. J. Climbing the Jaynes–Cummings ladder by photon counting. *J. Nanophotonics* **6**, 061803–061803 (2012).
20. Fano, U. Effects of configuration interaction on intensities and phase shifts. *Phys. Rev.* **124**, 1866–1878 (1961).
21. Alsing, P., Guo, D. & Carmichael, H. J. Dynamic stark effect for the Jaynes-Cummings system. *Phys. Rev. A* **45**, 5135–5143 (1992).
22. Fushman, I. *et al.* Controlled phase shifts with a single quantum dot. *Science* **320**, 769–772 (2008).
23. Wilkens, M. & Rzaewski, K. Resonance fluorescence of an arbitrarily driven two-level atom. *Phys. Rev. A* **40**, 3164–3178 (1989).
24. Müller, K. *et al.* Ultrafast Polariton-Phonon dynamics of strongly coupled quantum Dot-Nanocavity systems. *Phys. Rev. X* **5**, 031006 (2015).
25. Roy, C. & Hughes, S. Polaron master equation theory of the quantum-dot mollow triplet in a semiconductor cavity-QED system. *Phys. Rev. B Condens. Matter* **85**, 115309 (2012).
26. Cohen-Tannoudji, C. *Atom-Photon Interactions* (Wiley, 2004).
27. López Carreño, J. C., Sánchez Muñoz, C., Sanvitto, D., del Valle, E. & Laussy, F. P. Exciting polaritons with quantum light (2015). 1505.07823.
28. Sánchez Muñoz, C. *et al.* Emitters of n-photon bundles. *Nat. Photonics* (2014).
29. Yu, Y. *et al.* Fano resonance control in a photonic crystal structure and its application to ultrafast switching. *Appl. Phys. Lett.* **105**, 061117 (2014).
30. Johansson, J., Nation, P. & Nori, F. Qutip: An open-source python framework for the dynamics of open quantum systems. *Computer Physics Communications* **183**, 1760 – 1772 (2012). URL http://www.sciencedirect.com/science/article/pii/S0010465512000835.

### Acknowledgements

We gratefully acknowledge financial support from the Air Force Office of Scientific Research, MURI Center for Multifunctional light-matter interfaces based on atoms and solids and support from the Army Research Office (grant number W911NF1310309). KAF acknowledges support from the Lu Stanford Graduate Fellowship and the National Defense Science and Engineering Graduate Fellowship. KM acknowledges support from the Alexander von Humboldt Foundation. AYP acknowledges support from the Bechtolsheim Stanford Graduate Fellowship. YAK acknowledges support from the Stanford Graduate Fellowship and the National Defense Science and Engineering Graduate Fellowship. KGL acknowledges support from the Swiss National Science Foundation.

### Author contributions

KAF and KM conceived and performed the experiments. KF performed the theoretical work and modelling. AR fabricated the device. TS performed MBE growth of the QD structure. AP, YAK, CD, and KGL provided expertise. JV supervised the project. All authors participated in the discussion and understanding of the results.

### Additional information

Correspondence and requests for materials should be addressed to K.F. and J.V.

### Competing financial interests

The authors declare no competing financial interests.

# Supplementary information: Self-homodyne measurement of a dynamic Mollow triplet in the solid state

Kevin A. Fischer[1†⋆], Kai Müller[1†], Armand Rundquist[1], Tomas Sarmiento[1], Alexander Y. Piggott[1], Yousif Kelaita[1], Constantin Dory[1], Konstantinos G. Lagoudakis[1], Jelena Vučković[1⋆]

**Temporal and spectral filtering**

In equation (1) of the main text, we presented the ideal spectrum as measured by an infinite-bandwidth detector. However, the experimental spectra were additionally filtered by our grating spectrometer's finite integration time and approximately Gaussian bandwidth of $\Gamma_{\text{FWHM}} = 16.5\,\text{pm}$. Although in principle a consequence of simultaneous temporal and spectral resolution is energy redistribution in time,[1] the spectrometer integrates on timescales extremely long compared to the correlation times. Thus, we are justified to model the measured spectra simply as the convolution of this Gaussian filter function with the ideal spectra (as given by equation [1]). As a result, all of our quantum-optical simulations and fits have been convolved with the spectrometer's roughly Gaussian response.

**Broadband compared to pulsed transmission measurements**

In the main-text, the spectra presented in Figs 1c and 1d were taken under excitation from a broadband source of $\approx 15\,\text{nW/nm}$, while the spectra in Figs 2 and 3 were taken under excitation from a pulsed laser of FWHM $\tau_{\text{p}} = 100\,\text{ps}$. Here in Figs S1a and S1b, we additionally present transmission spectra where the pulsed excitation beam (40 nW) has been spectrally scanned across the system's resonances. In contrast to the broadband experiments (reproduced in Figs S1c and S1d) where the colorful excitation allowed collection of a single spectrum, the pulsed experiments required the total transmitted power to be monitored for each excitation frequency.

We now show that the Fano-like interference affects the transmission spectra under both excitation conditions in almost identical ways. Initially, we adjusted our system to minimize continuum-mode scattering. Spectra from Figs S1a and S1c were taken under this condition and both yield the typical JC resonances. Next, the focus of our sample was altered to bring in small contributions from the previously rejected continuum-mode scat-tering.[2,3] Then, spectra from Figs S1b and S1d were taken at this new sample focus condition and both yield the self-homodyne JC Fano-like resonances. In both cases, the Fano-like resonances result in a clear suppression of the coherently scattered light at the QD-like polariton's emission frequency.

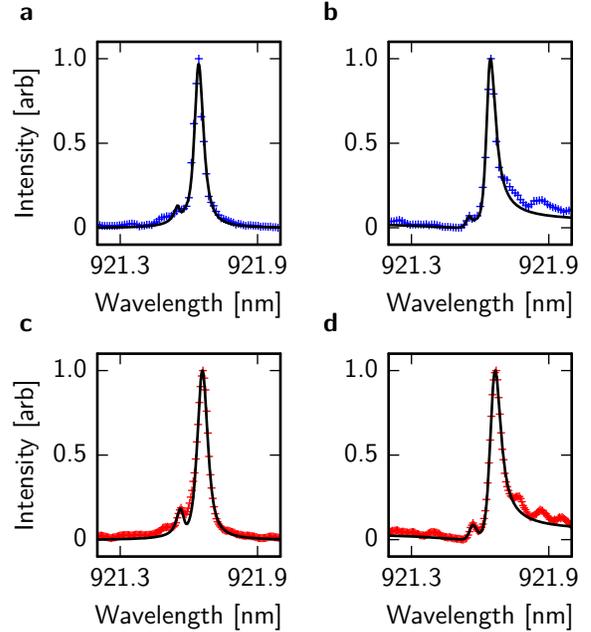

**Figure S1 | Comparison between continuous-wave and pulsed transmission through the strongly-coupled system.** Transmission spectra of the coupled quantum dot-cavity system taken at a detuning of $\Delta_{\sigma a} = -85\,\text{pm}$. **a,b,** Under excitation by a $\tau_{\text{p}} = 100\,\text{ps}$ FWHM pulsed laser at low power (40 nW), showing QD-like polaritonic emission on top of (**a**) a Lorentzian resonance and (**b**) a Fano resonance (for different focus conditions). Blue hashes indicate experimental data, black curves indicate quantum-optical fits. **c,d,** Under excitation by a continuous-wave broadband diode ($\approx 15\,\text{nW/nm}$), showing QD-like polaritonic emission on top of (**c**) a Lorentzian resonance and (**d**) a Fano resonance (taken at the same focus conditions as for [a] and [b], respectively). Red hashes indicate experimental data, black curves indicate quantum-optical fits.

Discrepancies between our quantum-optical model and the experiments arise primarily due to modulations in the cross-polarised suppression ratio resulting from imperfections in the table optics. The true features thus follow along the minima of these oscillations. Significant further error is introduced in Figs S1a and S1b by the precision of our pulse shaper when performing the excitation scans required for the transmission experiments. Additionally, we have only subtracted off the CCD dark counts in all experimental spectra. Hence, the relative decrease in the amplitude of emission at the QD-like polaritonic emission frequency (c.f. at 921.69 nm between Figs S1a and S1b or Figs S1c and S1d) is solely due to self-homodyne suppression from the Fano-induced interference.

---

[1]E. L. Ginzton Laboratory, Stanford University, Stanford, California 94305, USA
†These authors contributed equally
⋆email: kevinf@stanford.edu; jela@stanford.edu

## A simplified model of self-homodyne Fano suppression

Here we clarify the origin of our 95 % suppression metric in regard to the broadband spectra shown in the main text (Figs 1c and 1d). While the suppression of the total amplitude at the QD-like polariton emission frequency is on the order of a factor of two, this is not the important quantity for the self-homodyne measurements. Without suppression the signal consists of an incoherent and a coherent component. The Fano resonance suppresses the coherent component such that the signal consists of the same incoherent part but with a strongly suppressed coherent component.

To illustrate this point, in Figs S2a and S2b the spectra have been approximately decomposed into two components: a Lorentzian resonance representing the quantum mechanical influence of the QD (as the green dashed lines) and either a Lorentzian (Fig. S2a) or Fano (Fig. S2b) resonance representing coherent scattering from a cavity-like resonance (Fig. S2a, as the blue dashed line) or the cavity-like resonance and continuum modes (Fig. S2b, as the blue dashed line), respectively. The experimental data are represented by red hashes and the total fits (from the sum of each two components) are represented by the black lines. Now, consider the grey regions of interest that highlight emission at the QD-like polaritonic wavelength. In Fig. S2a, we can easily see that the Lorentzian tail of the cavity-like resonance significantly obscures incoherent emission from the QD. On the other hand, in Fig. S2b, the continuum modes are tailored to interfere with the cavity-like mode to generate a Fano resonance[4] of $q$ factor 5 (through careful adjustment of the excitation polarization and focus[3]). As a result, there is almost no overlap of the coherent emission with the incoherent emission in the region highlighted by the grey box. Thus, the suppression discussed is not in the overall amplitude at the QD frequency, but instead in the coherent part of the emission alone.

Comparing the relative amplitudes of the coherent components between Figs S2a and S2b results in a suppression of > 95 % of the coherent scattering at the QD-like polaritonic wavelength (see insets). This suppression is nearly identical with the experimental bound derived from comparing the net amplitudes of scattered light in Figs 2a and 2c (as discussed in the main text). We note that for a broadband spectrum obtained with low power densities (as presented Fig. S1) this effect might seem small. However, for a strong resonant excitation at this wavelength (as in Figs 2 and 3 of the main text) it is more easily visible due to saturation of the QD scattering compared to the cavity scattering.

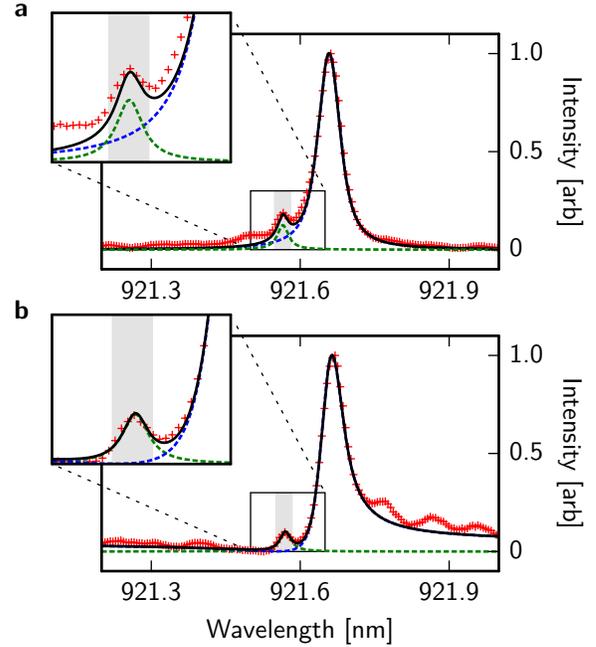

**Figure S2 | An approximate decomposition of the broadband transmission spectra.** Red hashes indicate the same spectra as in Figs S1c and S1d. **a,** Typical JC scattering: decomposition into first Lorentzian resonance (green dashed line represents quantum mechanical influence of the QD) and second Lorentzian resonance (blue dashed line represents coherent scattering from a cavity-like resonance). **b,** Self-homodyne JC scattering: Decomposition into Lorentzian resonance (green dashed line represents quantum mechanical influence of the QD) and Fano resonance (blue dashed line represents coherent scattering from the cavity-like resonance and continuum modes). Regions of interest for the self-homodyne measurements in the main text are highlighted by the grey boxes. Note how the Fano resonance suppresses the coherent scattering (blue dashed line) by >95 % in the grey region of interest.

## A complete model for self-homodyne Fano suppression

Although the simplified model discussed above very well captures the basic idea behind the self-homodyne Fano suppression, it is not technically complete. For a technically complete description, we apply the formal theory of quantum mechanical scattering discussed in the main text to the quantum optical fits of Figs 1c and 1d. We have decomposed these fits into their incoherently (green dashed lines) and coherently (blue dashed lines) scattered portions and present these results in Fig. S3. Considering the grey region of interest, we see that just as in Fig. S2, the Fano interference strongly suppresses the coherently scattered light from the cavity-like emission in order to reveal the quantum fluctuations due to the embedded quantum nonlinearity. In reality, we see that the effect of the QD-cavity strong coupling is to also induce incoherent scattering at the cavity-like polariton's wavelength. Additionally, for low powers the QD-like polariton does have some coherently scattered component. However, we have checked in simulation that these effects are minimal and that the toy model discussed above captures all of the relevant concepts for the self-homodyne measurements of the dynamic Mollow triplet presented in the main text.

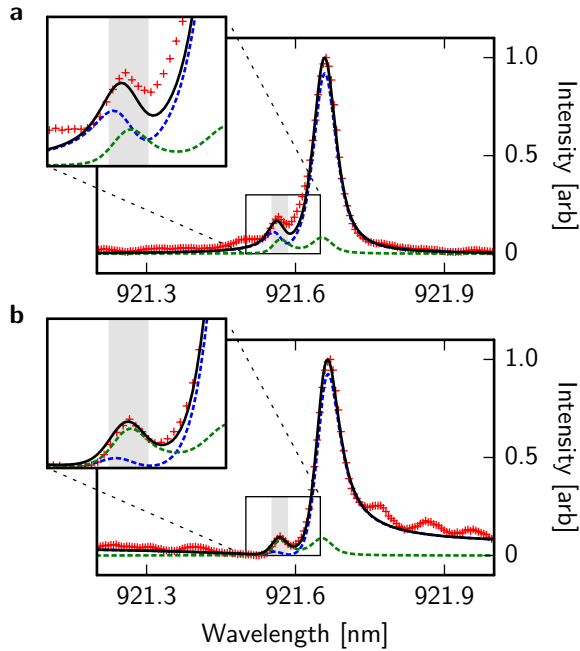

**Figure S3 | A quantum-mechanically complete decomposition of the broadband transmission spectra.** Red hashes indicate the same spectra as in Figs S1c and S1d. **a,b,** Quantum optical fits of the transmission spectra from Figs 1c and 1d in the main text are quantum-mechanically decomposed into their incoherently (green dashed lines) and coherently (blue dashed lines) scattered light. Decompositions are presented for (**a**) typical JC scattering and (**b**) self-homodyne JC scattering. Note the close similarity to the toy-model decompositions presented in Fig. S2, especially within the grey regions of interest.

## Power dependence of the Dynamic Mollow triplet

First, we discuss the triplet power-dependence for a traditional Mollow triplet. When a prototypical two-level system is strongly driven by a continuous-wave laser, its fluorescence is amplitude-modulated by the induced Rabi oscillations. If the laser resonantly drives a two-level system with a driving strength of $\Omega_{QD}$, then these single-photon oscillations occur at the Rabi frequency, leading to sidebands at frequencies of $\pm\Omega_{QD}$. Because the driving strength is proportional to the laser field, the sideband splitting follows the square root of the laser power.

Now consider our detuned CQED system, where the laser field almost exclusively couples to the cavity. While the lowest order QD-like polariton does indeed form a two-level system, its coupling to the laser field occurs indirectly — through coupling to the cavity. Because this energy exchange occurs in the strong coupling regime and for relatively mild detunings, effects of the cavity quantization should be apparent. For this reason, it is not immediately obvious that Mollow-like sidebands would even occur.

Here, we apply a simplified analysis to calculate the power-dependence of the CQED Mollow sidebands we observed. To begin, we analyse the case where the laser is resonant with the bare QD wavelength. First, we truncate the cavity-QD state basis to the three energetically lowest bare states (in the rotating frame of the laser frequency). From here, we can adiabatically eliminate the highest energy state to derive the dressed state splitting of $\pm 2g\Omega_a/\Delta_{\sigma a}$, where $\Omega_a$ is the cavity-laser coupling strength. In this frame, we see that the cavity and laser couplings act in concert to drive two-photon Rabi oscillations between the excited and ground states of the QD. Thus, these oscillations also give rise to a splitting that follows the square root of the laser power.

However, our CQED system is driven by a pulse that is twice the QD-like polaritonic state lifetime and hence at these timescales it is quite surprising to see the emergence of triplets at all. One might naively expect the system to simply emit a "smeared" version of the triplet with no clearly identifiable peaks. Instead, a recent theoretical analysis has shown that a complex phase interference effect lends these peaks definition for short pulses. As a result, the system produces modulation at slightly less than the peak Rabi frequency.[5] Based on these results, we adjusted the splittings with a prefactor to $\pm 0.8 \cdot 2g\Omega_a/\Delta_{\sigma a}$, which is reasonable in the context of the work of Moelbjerg *et al.*

## Comparison between CQED and two-level system dynamic triplets

Comparing our cavity QED dynamic triplet to the theoretical simulations of a two-level system's dynamic triplet in the work of Moelbjerg *et al.*,[5] one may notice a distinctive lack of additional side peaks from the CQED system compared to the two-level system. The lack of the additional side bands is not a flaw in our measurement but rather a physical feature of the CQED being measured. We measure the resonance fluorescence spectrum of the quantum dot (QD) - cavity QED polariton which differs from that of the standard two level system (such as a bare QD) studied theoretically by Moelbjerg *et al.* While in Moelbjerg's work the additional side peaks have comparable amplitudes to the main side peaks, several factors related to our cavity QED system contribute to their absence in our experiment. First, in a cavity QED system the coupling to the cavity mode greatly affects the peak heights and strongly enhances the peak closest to the cav-ity. Second, the phonon coupling further emphasizes this disparity and also generates a background of incoherent emission, among which the additional minor peaks sit.

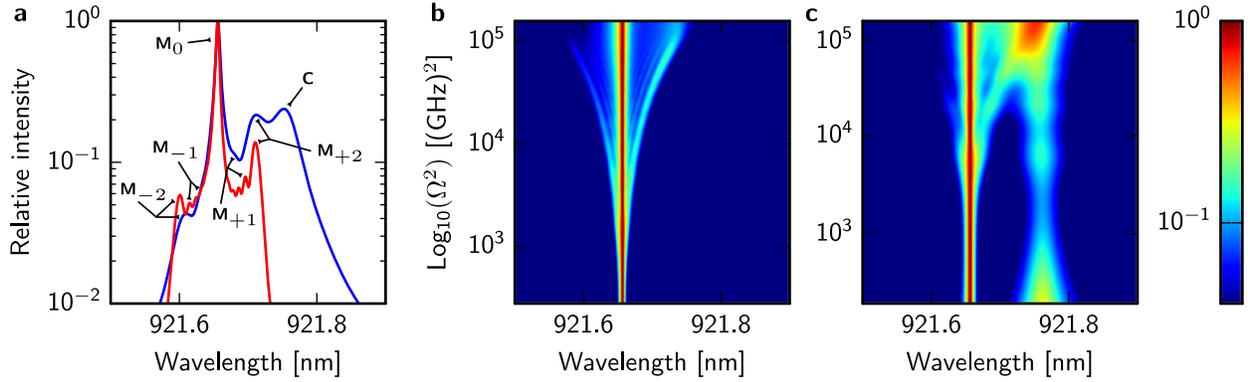

**Figure S4 | Theoretical analysis of the CQED dynamic triplet with and without phonon dephasing. a,** Line scans of the cavity mode operator incoherent spectra taken at a cavity driving strength of $\Omega = 2\pi \cdot 37$ GHz, are shown for both no phonon interaction (red) and with phonon interaction (blue). The peaks are labelled for clear references in the text. **b,c,** Stacked power incoherent spectra showing emergence of multipeak structure without (**b**) and with (**c**) phonon dephasing. Notice the dramatic asymmetry and smearing out of the peaks with phonon dephasing. Note: the stacked spectra are normalized to better visualize the emergence of the side peaks.

Before we continue our comparison, we want to firmly establish to the reader that our exploration of the CQED dynamic Mollow triplet was performed in exactly the same regime as the one studied theoretically[5] by Moelbjerg *et al.* In particular, we drive the system with pulses that are only 2.2 x longer than the state lifetime (as confirmed by our measurements of both the pulse length and the polariton lifetime[6]). This is the first demonstration of the solid-state Mollow triplet in the dynamic regime studied theoretically by Moelbjerg *et al.*, and thus any triplet we observe is indeed some form of a dynamic Mollow triplet.

Now we present a more detailed analysis, supported by simulation results, that shows the absence of the additional side peaks for a dynamic Mollow triplet in the CQED regime that we study. In Fig. S4, we compare simulations without and with phonon interaction (and excluding experimental limitations of the finite spectral resolution). The simulations without/with phonon interaction are shown as the red/blue line in Fig. S4a and as the colorplot in Fig. S4b/c. For the simulations without phonons, we increased pulse length so that the pulse-length to polariton lifetime ratio remained constant.

First, compare our Fig. S4a to Moelbjerg's Fig. 1a: the red trace (without phonon interaction) clearly has the same number of side peaks, but the peak amplitudes are radically different. In Moelbjerg's work, all side peaks have roughly the same amplitude and are approximately two orders of magnitude weaker than the central peak. However, the CQED side peaks are an order of magnitude stronger to begin with. Moreover, they show a striking asymmetry due to the strong quantum dot - cavity interaction. The peaks spectrally closer to the cavity wavelength (denoted as C in Fig. S4a) more readily mix with the cavity states and hence the strongest peak occurs nearest to the cavity (the $M_{+2}$ peak). Meanwhile, the next strongest peak (the $M_{+1}$ peak) is already weaker due to this effect. Note that the $M_{\pm 2}$ peaks most closely resemble the standard Mollow peaks.

Like in Moelbjerg's Fig. 1b, a colorplot showing the emergence of this multipeak structure is shown in Fig. S4 — again, the strong asymmetry due to a uniquely CQED effect can already be observed. The addition of phonon coupling, however, dramatically emphasizes this asymmetry. Comparing the red and blue traces in Fig. S4a shows that the $M_{+2}$ peak has been significantly enhanced, but at the expense of the $M_{-2}$ and $M_{+1}$ peaks. We also note that the cavity wavelength now shows emission due to the phonon assisted transfer process.[6] Strikingly, phonon dephasing in Moelbjerg's work on bulk quantum dots simply destroys the entire multipeak structure at the temperatures we performed our measurements (roughly 30 K). In contrast, the phonon dephasing in the CQED dynamic triplet emphasizes the $M_{\pm 2}$ peaks at the expense of the $M_{\pm 1}$ peaks. Whatever extremely weak evidence of additional side peaks that might be interpreted into Fig. S4c is washed out by experimental noise and an experimentally limited spectral resolution.

**References**

1. Eberly, J. H., Kunasz, C. V. & Wodkiewicz, K. Time-dependent spectrum of resonance fluorescence. *J. Phys. B At. Mol. Opt. Phys.* **13**, 217 (1980).
2. Galli, M. *et al.* Light scattering and fano resonances in high-q photonic crystal nanocavities. *Appl. Phys. Lett.* **94**, 071101 (2009).
3. Vasco, J. P., Vinck-Posada, H., Valentim, P. T. & Guimãraes, P. S. S. Modeling of fano resonances in the reflectivity of photonic crystal cavities with finite spot size excitation. *Opt. Express* **21**, 31336–31346 (2013).
4. Fano, U. Effects of configuration interaction on intensities and phase shifts. *Phys. Rev.* **124**, 1866–1878 (1961).
5. Moelbjerg, A., Kaer, P., Lorke, M. & Mørk, J. Resonance fluorescence from semiconductor quantum dots: beyond the mollow triplet. *Phys. Rev. Lett.* **108**, 017401 (2012).
6. Müller, K. *et al.* Ultrafast Polariton-Phonon dynamics of strongly coupled quantum Dot-Nanocavity systems. *Phys. Rev. X* **5**, 031006 (2015).